\documentstyle[aps,preprint]{revtex}

\begin{document}

\draft

\title{
Self-Organized Criticality and Synchronization in a Lattice
Model of Integrate-and-Fire Oscillators
}

\author{
\'{A}lvaro Corral, Conrad J. P\'{e}rez,
Albert D\'{\i}az-Guilera, and Alex Arenas
}
\address{
Departament de F\'{\i}sica Fonamental, Facultat de
F\'{\i}sica,\\ Universitat de Barcelona, Diagonal 647, E-08028
Barcelona, Spain
}

\maketitle

\begin{abstract}
We introduce two coupled map lattice models with nonconservative
interactions and a continuous nonlinear driving. Depending on
both the degree of conservation and the convexity of the driving
we find different behaviors, ranging from self-organized
criticality, in the sense that the distribution of events
(avalanches) obeys a power law, to a macroscopic synchronization
of the population of oscillators, with avalanches of the size of
the system.
\end{abstract}

\narrowtext

     A few years ago, Bak, Tang, and Wiesenfeld \cite{BTW} coined
the term self-organized criticality (SOC) to describe the
phenomenon observed in a particular cellular automaton model,
nowadays
known as the sandpile model. This system is critical in analogy
with classical equilibrium critical phenomena, where neither
characteristic time nor length scales exist. However, in SOC one
deals with dynamical nonequilibrium statistical properties and
the system evolves naturally to the critical state without any
tuning of external parameters.

     Several cellular automata models exhibiting SOC have been
reported in the literature. In the original sandpile model of
Bak {\em et al.} \cite{BTW}, the system is perturbed externally
by a random addition of sand grains. Once the slope between two
contiguous cells has reached a threshold value, a fixed amount of
sand is transferred to its neighbors generating a chain reaction or
avalanche. The critical state is characterized by a power-law
distribution of avalanche sizes, where the size is the total number
of toppling events. Taking this model as a reference, different
dynamical rules have been investigated leading to a wide variety of
universality classes.

     In the original noise driven models \cite{BTW,Zhang} it was
shown that conservative interaction rules were crucial to obtain
SOC \cite{MKK} but more recently the interest has been focused
on systems with a continuous driving such as stick-slip (SS)
models of earthquakes \cite{BK,FF,OFC,COB,OC,SGJ,RK,Grass} which
exhibit SOC without a conservation law.  The first of these
models was studied by Feder and Feder (FF) \cite{FF}; when
averaging over different samples a power-law distribution of
avalanche sizes is observed, but simple realizations exhibit a
periodical behavior that depends on the initial configuration.
Furthermore, the role played by nonconservation is unclear in
terms of the redistribution of forces after a slip process.
Lately, however, it was shown that deterministically driven models
with other interaction rules do exhibit SOC for different levels of
conservation \cite{OFC,COB,SGJ,Grass}.

All these SS models have strong analogies with certain models of
integrate-and-fire oscillators (IFO), widely analyzed to study the
behavior of populations of pacemaker cells and other biological
systems \cite{Peskin,Winfree,Ermen,SRKTMW,Abbott,MS,Kura,HSS}.
Of particular interest to us is the resemblance with a model
discussed by Mirollo and Strogatz (MS) \cite{MS}, who considered a
large assembly of IFO to show that under certain conditions the
stationary state presents perfect synchrony among all the elements of
the population. These conditions are: (i) a nonlinear convex driving
for the individual dynamics of each unit and (ii) long-range
interactions between them. With a linear driving it is possible to
have some temporal coherence -entrainment- but not to ensure that the
whole assembly will be synchronized. This study highlights the role
of the driving mechanism on the final state of the system. In the
SS and MS models there is an intrinsic dynamics -the driving-
leading the elements of the system to the threshold.  When a block
(oscillator) reaches the threshold it slips (fires) and this produces
a change in the state of its neighbors. This process can produce
further slips (firings) generating an avalanche. During the
propagation of an avalanche, the natural dynamics is stopped and the
collective behavior is determined by some interaction rules. When the
avalanche is completed, all the elements are below the threshold and
the driving governs again the dynamics of the system. Notice that
while in SOC models one studies the avalanche size distribution, in a
model of IFO one describes the state of the system after the end of
the interactive process, looking, for instance, at the level of
mutual entrainment between units. Essentially both models contain
the same basic ingredients and from this point of view SOC and
synchronization might be considered as two sides of the same
coin.

     In view of the analogy between models displaying SOC and IFO
models showing complex patterns of synchronization, the purpose of
this Letter is to develop a general framework where a diversity of
macroscopic behaviors can be observed by an appropriate choice of the
parameters that describe the dynamics of the elementary units as well
as the interaction rules among them. Keeping this goal in mind, we
are going to analyze the effect of a nonlinear convex driving in
models that do display SOC with a linear driving \cite{OFC,foot.SGJ},
and to observe under which conditions SOC breaks down and relaxation
oscillations show up.

     We study a coupled map model on a two-dimensional square
lattice of size $L \times L$ with open boundary conditions.
Each cell is characterized by two variables: a phase variable
$\varphi$, that increases linearly with time, and a state variable
$E$, which we call energy but can have different physical
interpretations. They evolve in time as \cite{Peskin}
\begin{mathletters}
\begin{equation}
d \varphi /dt=1
\end{equation}
\begin{equation}
dE/dt = \gamma (K-E)
\end{equation}
\label{fidete}
\end{mathletters}
with $K=1/(1-e^{- \gamma })$. Assuming $E(\varphi=0)=0$ and
$E(\varphi=1)=1$, these variables are related through the following
relations \cite{MS}
\begin{mathletters}
\label{peskin}
\begin{equation}
E( \varphi )=K \left( 1-e^{-\gamma  \varphi } \right)
\end{equation}
\begin{equation}
\varphi (E)=\frac{1}{\gamma}\ln \frac{K}{K-E}
\end{equation}
\end{mathletters}
The cycle period has been taken to be one without loss of generality
and $ \gamma $ is a measure of the convexity of the driving.  From
Eq. (\ref{fidete}) we can see that only when $\gamma =0$ (linear
driving) the energy of the cells will increase uniformly with time.

     Once a cell becomes critical ($E_{i,j} \geq E_c=1$) it
"fires" and transfers energy to the four nearest neighboring cells
according to the following rules \cite{Zhang,OFC}:
\begin{mathletters}
\label{rules}
\begin{eqnarray}
E_{i \pm 1,j} &\rightarrow & E_{i \pm 1,j} + \epsilon E_{i,j}
\nonumber \\
E_{i,j \pm 1} &\rightarrow & E_{i,j \pm 1} + \epsilon E_{i,j}
\label{zhang}\\
E_{i,j} &\rightarrow & 0,\nonumber
\end{eqnarray}
where $ \epsilon $ is the dissipative rate and $ \epsilon
=0.25$ corresponds to a conservative dynamics.  This model
differs from FF and MS where the interaction rules are
\begin{eqnarray}
E_{i \pm 1,j} &\rightarrow & E_{i \pm 1,j} + \epsilon
\nonumber \\
E_{i,j \pm 1} &\rightarrow & E_{i,j \pm 1} +  \epsilon
\label{ff}\\
E_{i,j} &\rightarrow & 0. \nonumber
\end{eqnarray}
\end{mathletters}
In this case the system is not conservative for any value of $
\epsilon $. These rules are a short-range version of the
infinite-ranged rules used in Ref. \cite{MS}. Notice that the concept
of absorption is removed from the MS model, so that when a cell
becomes critical either by its own dynamics or by the
interactive process, it always transfers energy to (modifies the
phase of) its neighbors. The energy of a cell may be larger than
$E_c=1$ but this property is not inconsistent with Eq. (\ref{peskin})
since a given site may only have $E>1$ when an avalanche is
triggered, i.e., when the interaction rules control the dynamic
behavior of the system. However, there is an important difference
between the two interaction rules: in terms of IFO (\ref{zhang})
implies that the phase response of an oscillator that receives a
pulse not only depends on the current phase but also on the energy of
the element that has fired, while for (\ref{ff}) the energy of the
firing cell is irrelevant.

     Starting with a random distribution of phases, we let the
phase of the cells to evolve according to Eq. (\ref{fidete}),
until one of them reaches the threshold value and energy is
redistributed following (\ref{rules}).  We have introduced a
certain amount of noise to ensure that the driving makes only
one cell to fire. If several cells reach the threshold
simultaneously the noise discriminates between them by choosing
one at random. In this way it is ensured that two avalanches cannot
overlap. Once all cells have an energy below the threshold, the
system is driven again. This dynamics involves two  time scales, one
for the intrinsic dynamics of the units and
another for the interactions; in SS models of earthquakes the
first scale corresponds to the motion of the tectonic plates and
the second one to the duration of the earthquakes; since the
former is orders of magnitude larger than the second one, we are
going to assume that the avalanches are instantaneous. This
assumption has also been made in recent studies on IFO
\cite{MS,CA}, but in order to discuss more realistic models it is
necessary to take into account some further ingredients such as the
time that signals need to propagate through the lattice or the
refractory time associated to the response of a cell, just to mention
a few. Their effects are currently under study.

     We have performed numerical simulations for different
values of $\epsilon$ and $\gamma$ on lattice sizes up to $L=64$.
In Fig.~1 we have plotted the schematic phase diagram of model
(\ref{zhang}) in terms of these parameters. Three regions with a
clear different macroscopic behavior after $3 \cdot 10^{6}$
avalanches are observed. The features of each region are discussed in
the next paragraphs.

     In region A there is only one type of avalanche that sweeps
the whole system when it has reached the stationary state.  The
avalanche starts at a given cell and propagates forming a
diamond-shaped front of firing cells due to the underlying
square lattice structure.  Each site fires exactly once and when
the avalanche is over the driving acts until a cell fires
again. This cell (seed) is always the same. Based on the
structure of the front we can describe analytically this
repetitive situation.  Once the seed has fired it gets four
firings from its neighbors and does not fire again.  Its nearest
neighbors get one firing from the seed, then fire, and then receive
three firings from their neighbors. In this way, all the cells get a
fixed number of firings (before and after reaching the
threshold) which only depends on their position on the lattice.
Once the avalanche is over, the system has a well defined
distribution of energies which depends on the interaction rules: for
(\ref{ff}) the seed is at $4 \epsilon $, the cells which form the
vertices of the front are at $3 \epsilon $, the boundaries at
$\epsilon $, the corners have zero energy, and the remaining cells
(the bulk) at $2 \epsilon$. Now, all the cells increase their
respective phases uniformly, the seed being the one that reaches the
threshold first. The avalanche must be able to reach the zero energy
cells on the corners and enable them to fire. Thus the energy
corresponding to a phase increment of $1- \varphi (4 \epsilon)$ plus
the energy of two firings coming from their neighbors must be larger
than the threshold, i.e.,
\begin{equation}
E\left( 1- \varphi (4 \epsilon )\right) +2 \epsilon  \geq 1.
\end{equation}
A simple calculation yields
\begin{equation}
 \epsilon   \leq  \frac{2-K}{4}
\label{epsilon}
\end{equation}
which is the relation between $ \epsilon $ and $ \gamma $ that
must be satisfied for model (\ref{ff}) in order to show
relaxation oscillations involving all the cells. This relation
has been checked through simulations finding an excellent
agreement. Notice that after each avalanche the cells forming the
bulk have the same phase. Therefore, the system presents a
macroscopic synchronization among almost all the elements of the
lattice.

     Now, let us consider rule (\ref{zhang}) which gives SOC for
a linear driving. The main difference with respect to the above
situation is that we have to replace $ \epsilon $ by an
effective value $\overline{ \epsilon }$ since the energy of a
given site can be larger than 1 when the avalanche propagates
through the lattice. Within a mean-field approximation we will
assume that this energy is the same for each cell. For this
model, when the avalanche has finished, the seed has a phase $
\varphi (4\overline{ \epsilon })$ and to fire again it has to
increment it by an amount $1- \varphi (4\overline{ \epsilon })$.
Since its neighbors are at $ \varphi (3\overline{ \epsilon })$
the condition to repeat permanently this situation is
\begin{equation}
4\overline{ \epsilon } = 4  \epsilon  \left[ E \left(
\varphi (3\overline{ \epsilon })+1- \varphi (4\overline{
\epsilon }) \right) + \epsilon \right].
\label{epsilonbarra}
\end{equation}
This equality in addition with Eq. (\ref{epsilon}) (replacing
$\epsilon$ by $\overline{ \epsilon }$), give the following
relationship between $\epsilon$ and $\gamma $
\begin{equation}
\epsilon \leq \frac{1}{16}\left[\sqrt{K^2-52K+164} -
(K+6)\right].
\label{epsilondef}
\end{equation}
When this condition is fulfilled, only relaxation oscillations of
the size of the system can survive in the stationary state. The
curve corresponding to the equality has been plotted in Fig.~1
(solid line) and has been corroborated through simulations
(circles).  Each symbol is an average over 10 different random
initial distribution of phases.  The inset in Fig.~1 is an
example of the histogram of phases after an avalanche. The height of
the peak at $\varphi(2\overline{ \epsilon })$ scales as $L^2$, the
other two at $\varphi(\overline{ \epsilon })$ and at
$\varphi(3\overline{ \epsilon })$ scale as $L$, and finally there is
a finite number of cells that have zero phase and
$\varphi(4\overline{ \epsilon })$ whose height is negligible in this
plot.

     Thus we have a coupled map lattice model that exhibits SOC
when it is driven linearly and relaxation oscillations when the
nonlinear driving is sufficiently important. Now we focus on the
intermediate behavior (region B in Fig.~1). Starting in the
region with a macroscopic degree of synchronization we fix $
\gamma $ and increase $ \epsilon $. Slightly above the curve
given by Eq. (\ref{epsilondef}) an avalanche sweeping all the
lattice cannot repeat any more since the next one will be unable to
reach some of the cells in the boundaries and these cells will be the
starting point of future avalanches. This fact gives rise to a
periodic behavior with a discrete distribution of a few avalanche
sizes.  This is indeed what we have observed in the simulations; only
some values of avalanche sizes are present and those are very
sensitive to the initial random configuration. By increasing
$\epsilon$ the distribution of avalanche sizes, $D(s)$, varies from the
discrete distribution to a continuous one with some characteristic
lengths which can be identified by the peaks in Fig.~2; the peaks in
$D(s)$ scale with different powers of $L$ and a finite-size scaling is
not possible in this situation. Furthermore, this stationary
distribution does not depend on the initial conditions.

For larger values of $\epsilon$ (region C of Fig.~1), close to the
conservation line, the peaks disappear and we get a power law decay
with an exponential cutoff. In this case the distribution $D(s)$, for
different values of $L$, can be fitted into a single curve by a
proper finite-size scaling (see Fig.~3), which is the hallmark of
SOC. However, we have to stress that for large values of
$\gamma$ the power law behavior remains but it is followed by a
peak before the exponential cutoff. Nevertheless, the whole
distribution of avalanche sizes still presents a good finite-size
scaling. Another point to remark is that the exponents of the
finite-size scaling ($\alpha$ and $\beta$), and hence the slope of
$D(s)$ on a log-log scale, are continuous functions of $\gamma$ and
$\epsilon$; thus we can conclude that the nonlinear driving gives
rise to a broader spectrum of exponents than that studied in the
linear case \cite{OFC}. We have identified the transition from
regions B to C (squares in Fig.~1) by fixing $\gamma$ and increasing
$\epsilon$ up to the appearance of a finite-size scaling. The dashed
line is an exponential fit of the numerical results that we
extrapolate to the linear driving case ($\gamma = 0$) and for large
values of the convexity of the driving. The extrapolation to the
linear driving gives a value of $\epsilon \simeq 0.16$ below
which there would be no SOC, which is in agreement with recent
results by Grassberger \cite{Grass} on very large lattices.

     In summary we have studied two coupled map lattice models
where different behaviors, ranging from a power law decay of
the distribution of avalanche sizes and finite-size scaling,
characteristics of SOC, to relaxation oscillations
with a macroscopic degree of synchronization, can be observed.
This is achieved by an appropriate choice of the parameters
governing the dynamics of the elementary units and the
interaction rules between cells. Based on the spatial structure that
an avalanche sweeping the whole lattice has, we have found an
analytical relation between the convexity of the driving and the
level of conservation in the interaction rules that ensures that
such an avalanche will repeat continuously. This is the condition
that gives rise to relaxation oscillations.

     The authors are indebted to K. Christensen for very
fruitful discussions and for a critical reading of the
manuscript. This work has been supported by CICyT of the
Spanish Government, grant \#PB92-0863.

\begin{figure}
\caption[]{Schematic phase diagram in terms of $ \gamma $ (the
convexity of the driving) and $ \epsilon $ (the level of
conservation) for model (3a).  The symbols correspond to the
phase transitions observed in the simulations. For the B-C transition
the error bars denote that above them
we have always found SOC while below them there is no
power-law behavior. For the A-B transition the error bars
are given by the standard deviation over ten measures. The solid line
corresponds to our analytical result (\ref{epsilondef}) and the
dashed line is an exponential fit to the numerical data. The
inset displays the distribution of phases after an avalanche in
region A for a system of size $L=64$ with $\gamma=1.00$ and
$\epsilon=0.09$.}
\end{figure}

\begin{figure}
\caption{Distribution of avalanche sizes in region B.}
\end{figure}

\begin{figure}
\caption{Finite-size scaling ansatz of the distribution of
avalanche sizes in region C. In this case $\alpha=-2.9 \pm 0.1$ and
$\beta=2.0 \pm 0.1$}
\end{figure}


\begin{references}
\bibitem{foot.SGJ}
It is important to remark that in our case the nonlinearity is
introduced in the driving whereas in Ref. \cite{SGJ} the system
is driven uniformly and the
nonlinearity is introduced in the interaction rules.
\end{references}
\end{document}